\documentclass[aps,twocolumn,floatfix,superscriptaddress]{revtex4-1}
\usepackage{amsfonts,amsmath,amsthm,amssymb,amsopn,graphicx}
\usepackage{dsfont,bbm,xcolor}
\usepackage{enumitem}
\usepackage{hyperref}
\usepackage{breqn}
\usepackage{siunitx}
\usepackage{color}
\usepackage{braket}

\makeatletter
\let\cat@comma@active\@empty
\makeatother

\newcommand{\ii}{ {\rm i} }

\def\abs#1{\left | #1 \right |}

\newcommand{\ave}[1]{{\langle #1\rangle}}

\newcommand{\tr}{\rm{tr}}

\def\one{\mathbbm{1}}

\newcommand{\sinner}[2]{\langle \langle #1 |#2 \rangle \rangle}

\newcommand{\LL}{{\hat{\cal L}}}

\def\tr{{{\rm tr}}}

\def\one{\mathbbm{1}}

\begin{document}
\title{Exact bistability and time pseudo-crystallization of driven-dissipative fermionic lattices}

\author{Hadiseh Alaeian}
\affiliation{Elmore Family School of Electrical and Computer Engineering, Department of Physics and Astronomy, Purdue Quantum Science and Engineering Institute, Purdue University, West Lafayette, Indiana 47907-2035, USA}

\author{Berislav Bu\v{c}a}
\email{berislav.buca@physics.ox.ac.uk}
\affiliation{Clarendon Laboratory, University of Oxford, Parks Road, Oxford OX1 3PU, United Kingdom}

\date{\today}

\begin{abstract}
The existence of bistability in quantum optical systems remains a intensely debated open question beyond the mean-field approximation. Quantum fluctuations are finite-size corrections to the mean-field approximation used because the full exact solution is unobtainable. Usually, quantum fluctuations destroy the bistability present on the mean-field level. Here, by identifying and using exact modulated semi-local dynamical symmetries in a certain quantum optical models of driven-dissipative fermionic chains we exactly prove bistability in \emph{precisely} the quantum fluctuations. Surprisingly, rather than destroying bistability, the quantum fluctuations themselves exhibit bistability, even though it is absent on the mean-field level for our systems. Moreover, the models studied acquire additional \emph{thermodynamic} dynamical symmetries that imply persistent periodic oscillations in the quantum fluctuations, constituting pseudo-variants of boundary time crystals. Physically, these emergent operators correspond to finite-frequency and finite-momentum \emph{semi-local} Goldstone modes. Our work therefore provides to the best of our knowledge the first example of a provably bistable quantum optical system. 
\end{abstract}

\maketitle
\section{Introduction}~\label{sec:intro}
Bistability in driven-dissipative models usually means the presence of two possible stationary states of the system that can be distinguished by local observable measurements. Although on the level of the mean-field approximation it can be easily established whether or not it exists, its actual existence, in particular in low-dimensional strongly interacting systems, remains quite controversial with both theoretical and experimental work reporting differing conclusions~\cite{Ding2020,Ferri2021,Landa,Sarang,order,vaporization,Letscher,JuanJose,Parmee_2020,Bistability1,baas2004optical,Orazio2,bistability4,Piazza,TobiReview,iPEPO,lambert2021quantum,Bruder,Pizzi,Lesanovsky2}. The existing approaches usually rely on sophisticated theoretical techniques for including perturbations of finite-size corrections to the mean-field or large scale efficient numerical simulations such as t-DMRG \cite{JuanJose} or projected entangled pair states (PEPS) \cite{iPEPO}. The general lore in the literature is that the lower dimension - the more likely it is that the full quantum fluctuations (finite-system size corrections) will destroy the bistability and restore the generic unique stationary state of the model~\cite{owen2018quantum,bistability2,bistability3,Cabot}. However, in general, the question remains unsettled in any dimension (see e.g. \cite{iPEPO} for an advanced numerical study in two-dimensions). Therefore, exact results on this controversial problem are desirable for understanding self-organization in non-equilibrium quantum systems. 

In this paper we show a lower bound on the bistable finite-size fluctuations for a class of realistic driven-dissipative fermionic lattice models at certain driving values. Remarkably, in contrast to other potentially bistable models e.g.~\cite{owen2018quantum}, here, rather than being detrimental, quantum fluctuations are essential for bistability.

Our approach is based on identifying a novel modulated \cite{Pollmann} spectrum generating algebra (SGA) \cite{SGA}, which is semi-local and fermionic. Because standard dynamical symmetries \cite{Buca_2019} are extensive and local SGA \cite{Marko1,SanjayReview} and the SGA here is extensive and semi-local, we call it a semi-local \cite{Mauriziosemilocal} dynamical symmetry. This dynamical symmetry, being fermionic, cannot be relegated to a non-Abelian symmetry, unlike previously known cases based on closed algebras settling the question of whether such operator relations are possible~\cite{Marko2}. Later, for sake of simplicity, we specialize the general dissipative-driven fermionic model to a quadratic model and show that these models have an infinite set of emergent (thermodynamic) super-extensive raising operators that we call super-extensive dynamical symmetries and provide evidence that the total effect of all these operators is that the model displays very slow finite-size decay due to the presence of strong symmetries \cite{BucaProsen}, which guarantee degenerate stationary states (null space of Liouvillian). These have attracted lots of interest mainly due to their utility for quantum information storage (e.g.  \cite{AlbertJiang,Albert1,albert_thesis,Kollath,strongsymmetry1,strongsymmetry2,weaksymmetry2,strongsymmetry3,weaksymmetry,strongsymmetry4,strongsymmetry5,UedaEta}). Even though, in general, they are not necessary for the existence of bistability because degenerate stationary states may emerge in the thermodynamic limit only (without a strong symmetry), nor are they sufficient because all of the degenerate stationary states implied by the strong symmetries may have the same expectation values for local observables. However, in our case strong symmetries do guarantee bistability in local observables. 

There are two recently introduced concepts, dissipative time crystals~\cite{Buca_2019,Kessler2021,kongkhambut2022observation,Seibold,Jamir1,Seibold2,BucaJaksch2019,Cosme2,Fabrizio,Esslinger,Esslinger2,BucaJaksch2019,Booker_2020,Chinzei,Chinzei2,sarkar2021signatures}, which are systems that have persistent oscillations induced by the dissipation, and boundary time crystals \cite{Fazio,Shammah,Lesanovsky,boundary2,boundary4}, which have persistent oscillations in the thermodynamic limit only (cf. discrete, driven versions of time crystals under dissipation \cite{Ueda,drivendissipativeBH1,lledo2019driven,Mitra,LazaridesDissipation,mcginley2021absolutely,DisTC2021A} and other non-stationary phenomena beyond observables e.g. \cite{gameoflife,Entaglementresonance,yuan2022quantum,Carlos1,scars,scars1,scarsdynsym2,scarsdynsym1,scarsdynsym3,scarsdynsym6,ScarsNew1,Pakrouski,scars9,scarsNew2,Olalla1,Olalla2}). As the oscillations in our model are persistent in the thermodynamic limit, the model may be understood as a boundary time \emph{pseudo}-crystal, with \emph{pseudo-} implying that the oscillations amplitude decays with the system size for initial states with low entanglement, similarly to long-range order in a pseudo-condensate \cite{etapairing}. 

\section{Organization of the paper}
The paper is organized as follows: in Sec.~\ref{sec:model} we introduce the model which is an extension of a 1D Kitaev chain including pair-particle interactions. This model is related to quantum optical setups that are used to study bistability. There we introduce the strong dynamical symmetries of the conserved dynamics emerging as modulated non-local and semi-local fermionic and spin operators, respectively. Later, we generalize the results of closed system dynamics to explore new phases such as multi-stability and dissipative time crystalline phase that can be established in the open quantum system subject to some Hermitian jump operators. In Sec.~\ref{sec:TD and dynamical symmetries} we study the dynamical symmetries of such model in the most general form and in the thermodynamic limit, find a lower limit for the finite-size scaling in the system, and highlighting its relation to the well-known Goldstone modes. Section~\ref{sec: results} presents relevant numerical results of the non-interacting case showcasing the aforementioned phases via exact calculation of the Liouvillian spectrum and the dynamical evaluations of the correlations. Furthermore, we examine the finite-size scaling in such systems which supports the theoretical predictions of the previous sections. Finally, we conclude the paper in Sec.~\ref{sec:conclution} and discuss some future directions that can be pursued based on the results presented in this work.

\section{The model and semi-local dynamical symmetries}~\label{sec:model}

Consider the following interacting Kitaev chain model~\cite{Kitaev2003},

    \begin{eqnarray} 
    & H = -\frac{1}{2} \sum_{i=1}^N  \left(w c_i^\dagger c_{i+1} + w^* c_{i+1}^\dagger c_i \right)  + \Delta c_i c_{i+1} + \Delta^* c_{i+1}^\dagger c_i^\dagger \nonumber \\
    & + \mu \sum_{i}\left(n_i - \frac{1}{2}\right) + \sum_{m,K} V_{m,K} \prod_m^K (c_m + c_m^\dagger), \label{eq:Hamiltonian}
    \end{eqnarray}

with $c^\dagger_j$ and $c_j$ being the (Dirac) fermionic creation and annihilation operator, $w$ is the hopping amplitude, $\Delta$ is the p-wave pairing correlation, $\mu$ is the on-site chemical potential, $K$ is an even integer, and $V$ is a novel interacting (beyond-quadratic) term we here propose to model strong pairing correlations. The number operators are $n_j=c_j^\dagger c_j$ and the chain is subject to the periodic boundary condition, i.e. $c_{N+1} = c_1$. Here we will set $\mu=0$ and discuss generalizations in the next section.  

This model is commonly mapped onto a spin-$1/2$ transverse field Ising model at $V=0$~\cite{Fendley}. Note that the model we consider here is not mappable to integrable XYZ spin chains or non-interacting models in contrast to other interesting results \cite{Hosho2020,Chitov2018}. 

We first note that the following \emph{modulated} operator at momentum $k=\pi/2$,
\begin{equation}
A_0=\sum_{x=1}^N \exp(\ii \frac{\pi}{2} x ) (c_x+c_x^\dagger), \label{eq:Aoperator0}
\end{equation}
satisfies $[H(k),A_0(\pi/2)]=\omega_0 A_0 (\pi/2)$ with $\omega_0= \ii \Delta$ provided that $\Delta$ is purely imaginary~\footnote{for other $\Delta$ it does not fulfill this relation.} and mod $(N,4)=0$. $A_0$ is local in the fermionic basis, hence it is a fermionic dynamical symmetry of the model. It immediately implies that observables that have non-zero overlap with it can persistently oscillate \cite{Marko2}. We note that complex values of $\Delta$ may be physically obtained with e.g. constant phase gradients \cite{phasegradient1} or laser coupling as recently employed in the simulation of a bosonic ladder~\cite{Hung2021}.

We now define $m_j=\frac{1}{2}\one-n_j$, the parity operator $P_{j,k}=\prod_{q=j}^k m_q$, and $b_x=P_{1,x-1}c_xP_{x,N}$. 

Then a related non-local operator,
\begin{equation}
A=\sum_{x=1}^N \exp(\ii \frac{\pi}{2} x ) (b_x+b_x^\dagger), \label{Aoperator}
\end{equation}
likewise satisfies $[H(k),A(\pi/2)]=\omega A(\pi/2)$ with $\omega= \ii \Delta$, also at momentum $k = \pi/2$. 

Here, for reasons that will become apparent, we consider the standard Wigner-Jordan mapping~\cite{Jordan1928}, 
\begin{align}
&\tilde{P}_{j,k}=\prod_{x=j}^k \sigma^z_x \, , \nonumber\\ 
&c_j= \tilde{P}_{1,j-1} \sigma^-_j \, .\label{map}
\end{align}
Following the mapping in \eqref{map}, operator $A$ gets transformed to a \emph{semi-local} dynamical symmetry in the spin basis, i.e. its densities commute only with operators on one side,
\begin{equation}
\tilde{A}=\sum_{j=1}^N \exp(\ii \frac{\pi}{2} j ) \sigma^x_j \tilde{P}_{j+1,N}. \label{Aoperator1}
\end{equation}
Such semi-local \emph{symmetry} operators have been studied recently in the context of generalized hydrodynamic corrections in quadratic and integrable models where their existence was associated with the topological nature of the models~\cite{Mauriziosemilocal}. These new kinds of dynamical symmetries should be distinguished from both local extensive \cite{Marko1,Doyon1,Doyon2,Doyon3} and strictly local ones \cite{Thivan,OTOcrystal,Buca_2020} Topology likely plays a role in our model as it is intimately related to the Kitaev chain. We emphasize, that in the model studied here there is no obvious transformation that would allow mapping the semi-local dynamical symmetry into a semi-local non-Abelian symmetry while preserving the spatial locality of $H$.  

The operators $A$ and $A^\dagger$ satisfy fermionic anti-commutation relations $\{A,A^\dagger\}=\one$ and they are nilpotent $A^2=0$. 
It will be convenient to define Majorana fermions,
\begin{equation}~\label{eq:physical Majorana fermions}
\gamma_{2j-1} = c_j + c_j^\dagger \qquad \gamma_{2j} = i(c_j - c_j^\dagger).
\end{equation}
These fulfill the following anti-commutation relations,
\begin{equation} \label{eq:majoranaCAR}
\{\gamma_j,\gamma_m\}=2\delta_{jm}\, .
\end{equation}
The Hilbert-Schmidt inner product on the space of operators is defined as $\sinner{A}{B}=\tr({A^\dagger B})$, with a corresponding norm that we will use.

The existence of $A$ immediately implies the existence of a super-extensive (quadratic) charge $Q=A^\dagger A$, i.e. whose norm can be easily shown to grow as $\propto N^2$. In the Majorana basis it may be written as,
\begin{equation}
Q=\ii(\sum_{j=1}^{N} \ii^{j-1} \gamma_{2j})(\sum_{j=1}^{N/2} (-1)^{j} \gamma_{4j}) \, . 
\end{equation}
This \emph{Hermitian} operator defines a symmetry $S=e^{\ii Q}$ as $[H,S] = 0$ with two eigenvalues of $\pm 1$. 
The existence of $Q$ immediately implies memory of the initial state that decays as $1/N^2$ in local observables as quantified by infinite temperature auto-correlation functions via the Mazur bound. 

However, here we will study a dissipative model with an incoherent Markovian driving modeled by a Lindblad master equation,
\begin{equation} \label{eq:ddt}
    \frac{d\rho}{dt} = \LL[\rho].
\end{equation}
Here $\LL$ is a quantum Liouvillian of the form 
\begin{equation} \label{eq:masterequation}
     \LL[\rho] = -\ii [H,\rho]  + \sum_\mu \left( 2 L_\mu \rho L_\mu^\dag - \{L_\mu^\dag L_\mu, \rho\} \right),
\end{equation}
and the Lindblad jump operators are local incoherent dissipative driving terms of the form $L_j=\sqrt{\Gamma}(c^\dagger_j+c_j)=\sqrt{\Gamma} \gamma_{2j-1}$. The quadratic version of this model ($V_{m,K}=0$) has been precisely studied for its topological properties \cite{Moos}. 

In order to solve the dynamics of $\rho(t)$ it is useful to diagonalize $\LL$. Define $\lambda_k$ to be the eigenvalues of $\LL$ and $\rho_k, \sigma_k$ to be the corresponding right and left eigenoperators respectively,
\begin{equation}
    \begin{gathered}
    \LL [\rho_k] = \lambda_k \rho_k, \ \LL^\dag[\sigma_k] = \lambda_k^* \sigma_k, \\ \sinner{\sigma_{k}}{\rho_{k'}} = \delta_{k,k'}.
    \end{gathered} \label{eigensystem}
\end{equation}
Due to the semi-group properties of the Lindblad master equation all the eigenmodes are either stable or decaying, i.e. $\text{Re}(\lambda_k) \le 0$. They further always appear in complex conjugate pairs $\{\lambda_k,\lambda_k^*\}$. Since the jump operators $L_\mu$ are Hermitian in our model, the Lindblad equation is unital, and the identity matrix is a stationary state $\rho_0=\one$.

We are interested in the dynamics of observables $O(t)$ when we initalize the system in $\rho(0)$. Formally, the solution is
\begin{equation} \label{eq:formalsolution}
    \ave{O}(t) = \sum_k e^{t \lambda_k} \sinner{O}{\rho_k} \sinner{\sigma_k}{\rho(0)}\,.
\end{equation}
Purely imaginary eigenvalues are therefore necessary but not sufficient for persistent oscillations in physical observables due to possibly vanishing overlap with local observables, or initial states, or the presence of dense and incommensurate purely imaginary eigevalues (eigenfrequencies) ${\lambda_k}$ in the sum. 

As may be explicitly checked, the semi-local dynamical symmetry satisfies $[A,L_\mu]=[A^\dagger,L_\mu]=0$, $\forall \mu$ implying that it is a \emph{strong} dynamical symmetry of the dissipative model~\cite{buca2022algebraic,Carlos2}. This implies that the Lindblad master equation of the model has purely imaginary eigenvalues $\lambda_{\pm 1}=\pm \Delta$, for pure imaginary $\Delta$. Likewise, $S$ is a strong symmetry \cite{BucaProsen,Zhao}, which in turn implies that the non-equilibrium stationary state $\lambda_0=0$ is degenerate with at least dimension 2 coming from the two distinct eigenvalues of $S$. 

It is known that since the model is unital, $A$ is a strong dynamical symmetry and satisfies the fermionic canonical anticommuntation relations, that $\rho_1=A^\dagger$, $\rho_{-1}=A$, and the bi-orthogonal eigenmode in stationary state manifold is  $\rho'_{0}=\frac{1}{2}-A A^\dagger$ with eigenvalues $\lambda_{\pm1}=\Delta$ and $\lambda_0=0$. Likewise, due to unitality the left and right eigenmodes are each others conjugate transposes $\sigma_{\mp1}=\rho_{\pm1}^\dagger$. 

It is obvious that the presence of purely imaginary eigenvalues is not present in any local observable $O$ because non-local $\rho_{\pm1}$ do not have overlap with such observables. However, $\rho'_0$ does have overlap, but the overlap of the corresponding left eigenmode $\sigma'_0$ with the initial state $\rho(0)$ scales as $1/N$ for low-entangled initial states. This means that we may estimate in the long time limit for a local observable $O$,
\begin{equation}
O(t\to\infty) \propto \sinner{O}{\rho'_0} \sinner{\sigma'_0}{\rho(0)}\propto 1/N,
\end{equation} 
assuming, again low enough entanglement for $\rho(0)$, i.e. clustering of correlations (e.g. for product states) (note that we assume $\tr (O\rho_0)=\tr O=0$ for simplicity). 
Thus, the bistability of the model is visible in local observables only in the quantum fluctuations, i.e. as a finite-size effect $1/N$ beyond mean-field. The decay with system size will be also dictated by the overlap of the left stationary state with the initial state. 

Moreover, additional (dynamical) may emerge in the thermodynamic limit that will slow down the decay with the system size. We will study this in the next sections.

\section{Emergent semi-local dynamical symmetries in the thermodynamic limit}~\label{sec:TD and dynamical symmetries}

The origin of these semi-local dynamical symmetries may be understood by studying the quadratic version of the model in \eqref{eq:Hamiltonian} in the non-interacting limit, i.e. $H_0=H(V_{m,K}=0)$. The model is then the standard Kitaev pairing Hamiltonian that can be diagonalized with a Fourier transform followed by a Bogoliubov transformation. We obtain (up to an irrelevant shift),
%
%%%%%%%%%%%%%%%%%%%%%%%%%%%%%%%%%%%%%%%%%%%%%%%%%%%%%%%%%%%
\begin{figure*}[t]
    \includegraphics[width=1.0\linewidth]{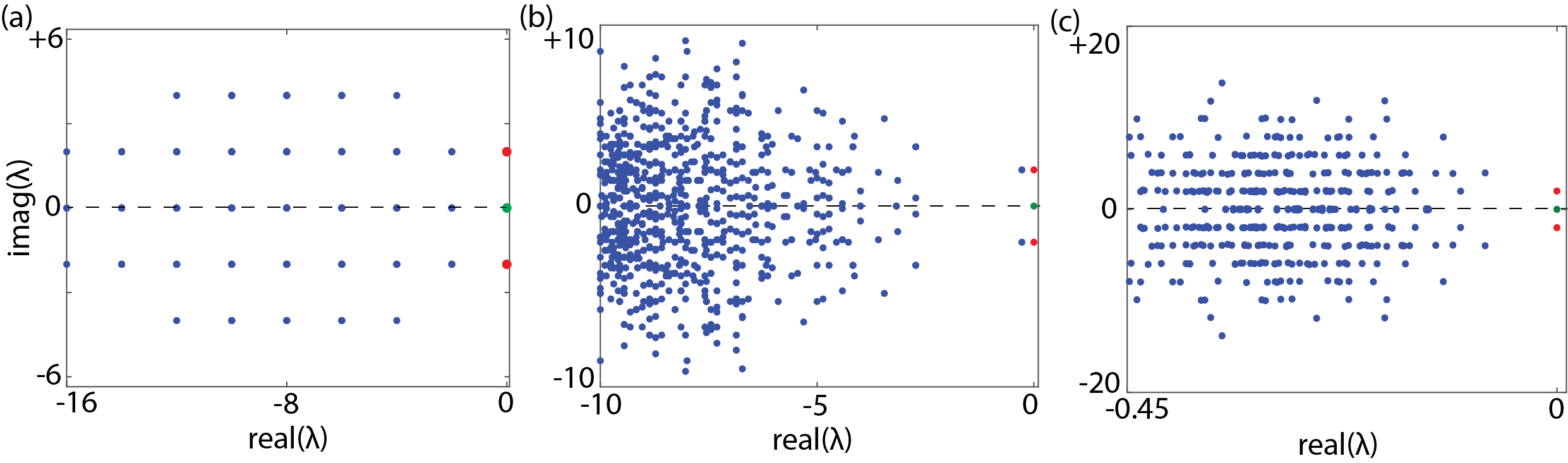}
    \centering
    \caption{~\label{Fig1} Liouvillian spectrum of non-interacting Kitaev chain for different chain length (a) N=4, (b) N = 12, and (c) N = 100 subject to periodic boundary conditions. In all cases $\mu = 0, w = 1, \Delta = i$, and red (green) dots show the pure imaginary (zero) eigenvalues.}
\end{figure*}
%%%%%%%%%%%%%%%%%%%%%%%%%%%%%%%%%%%%%%%%%%%%%%%%%%%%%%%%%%%%

\begin{equation}
H_0=\sum_k E_k d^\dagger(k) d(k),
\end{equation}
with the lowering operators of,
\begin{align}
& d(k)=u_k c(k) + v_k c^\dagger(-k)\, , \nonumber \\
& c(k) = \frac{e^{-\ii \frac{\pi}{4}}}{\sqrt{N}}\sum_{j=1}^N e^{-\ii k j}c_j, 
\end{align}
with $c(k)$ being the Fourier transform of the fermion annihilation operator at momentum $k$ and $d(k)$ being their Bogoliubov transformation where the (not-normalized) coefficients are defined as
\begin{align}
&u_k=-\frac{\ii \Delta  \sin (k) \sqrt{E_k-w \cos (k)-\mu }}{\sqrt{2} \left| \Delta  \sin (k)\right|  \sqrt{E_k}}\, , \nonumber \\
& v_k=\frac{\ii (E_k+w \cos (k)+\mu )}{\Delta \sin(k)} u_k, \label{eq:bcoeff}
\end{align}
and the energy is,
\begin{equation}
E_k=\sqrt{\left| \Delta  \sin (k)\right| ^2+(w \cos (k)+\mu )^2}\, .    
\end{equation}
The momentum is restricted to the first Brillouin zone $k=\frac{2\pi}{N} m$, $m=-\frac{N}{2}+1,\ldots,\frac{N}{2}-1,\frac{N}{2}$.

If for some $\kappa$, $u_\kappa=-v_\kappa$, we have up to a multiplicative constant,
\begin{equation}
d_\kappa=\sum_{j=1}^N e^{-\ii \kappa j} \gamma_{2j}\,. \label{eq:newdyn}
\end{equation}
Solving $u_k=-v_k$ using \eqref{eq:bcoeff} for purely imaginary $\Delta$ gives $\kappa=\cos^{-1}(-\frac{\mu}{w})$, which is a real momentum for $\abs{\mu}<\abs{w}$, coinciding with the topological phase of the Kitaev chain.  Since the interaction terms in $H$ contain only products of $\gamma_{2j-1}$, following the Majorana anti-commutation relations \eqref{eq:majoranaCAR}, the interaction term commutes with $d_\kappa$ in \eqref{eq:newdyn}. From this it directly follows that,
\begin{equation}
[H,d_\kappa]=-E_\kappa d_\kappa,
\end{equation}
where $E_\kappa=\left| \Delta  \sqrt{1 - \frac{\mu ^2}{w^2}}\right|$. 

Thus $d_\kappa$ is a modulated fermionic dynamical symmetry of the model. However, for general $\mu,w$ the dynamical symmetries exist only in the thermodynamic limit as for finite systems there is no solution for $\kappa$ for general $\mu,w$. In other words, they are emergent for systems that are large enough to have a continuum of momenta $k$ in the 1st B.Z. Hence, these dynamical symmetries are thermodynamically emergent.  For $\mu=0$ we get the original \eqref{eq:Aoperator0} from the previous section, obtained at $\kappa = \pi/2$, which is a valid solution provided that mod $(N,4)=0$.

We now note some useful identities. The Hamiltonian is parity-symmetric, i.e. $[H_0,P_{1,N}]=0$, and the Lindblad jump operators are parity-antisymmetric, i.e. $\{L_\mu,P_{1,N}\}=0$ and furthermore satisfy $\{L_\mu,d_\kappa\}=0$, as are the Majorana fermions in general $\{\gamma_j,P_{1,N}\}=0$. From this it follows directly that $[H_0,P_{1,N} d_\kappa]= - E_\kappa P_{1,N} d_\kappa$ and $[L_\mu,P_{1,N} d_\kappa]=0$ hence, $A_{\kappa}=P_{1,N} d_\kappa$ is a non-local strong dynamical symmetry. 

%%%%%%%%%%%%%%%%%%%%%%%%%%%%%%%%%%%%%%%%%%%%%%%%%%%%%%%%%%%
\begin{figure*}[t]
    \includegraphics[width=1.0\linewidth]{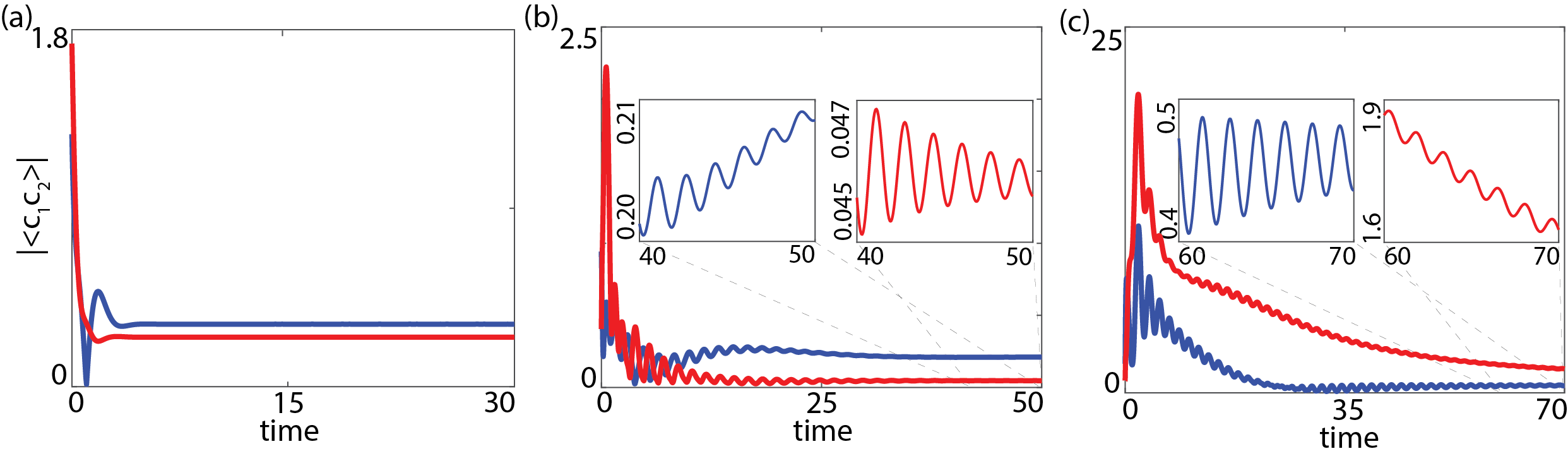}
    \centering
    \caption{~\label{Fig2} Time evolution of $|\braket{\hat{c}_1 \hat{c}_2}|$ showcasing the multi-stability for two randomized initial state and oscillatory behavior for (a) N=4, (b) N = 12, and (c) N = 100 subject to periodic boundary conditions.}
\end{figure*}
%%%%%%%%%%%%%%%%%%%%%%%%%%%%%%%%%%%%%%%%%%%%%%%%%%%%%%%%%%%%

Expanding for small $\varepsilon$ around $\kappa\pm\varepsilon$ we have $[L_\mu,P_{1,N}d_{\kappa\pm \varepsilon}]={\cal O}(\varepsilon)$. As $k\propto 1/N$, this implies that the dissipative gap (the real part of the eigenvalues of the Liouvillian) closes as $\propto 1/N$ into the same purely imaginary eigenvalues as the finite size Liouvillian. They are hence metastable \cite{metastability1,metastability2,metastability3}. We will confirm this in the next section in a concrete example. The closing of the Liouvillian gap is associated with algebraic (in time) relaxation of the dynamics, but as we will see in the next section, it can also lead to larger quantum fluctuations (i.e. slower decay of bistability with $N$). They are essentially similar to Goldstone modes, except they exist at finite frequency and momentum. Their presence leads to a different behaviour than the standard power-law (in time) decay of bistability and oscillations that are usually studied for closing Liouvillian gaps \cite{Gap1,Gap2,Gap3}

It is important to note that, even though $A_{\kappa}^2=0$, $A_{\kappa+\varepsilon}A_{\kappa}\neq0$ and thus these operators do have overlap with local operators (this follows from $P_{1,N}^2=\one$). This will lead to boundary pseudo-time crystal behavior in local observables, as we will study in the next section. 
\section{Results}~\label{sec: results}
From now on, for sake of simplicity, we will focus on the quadratic model, noting that the general conclusion, according to the discussion in the previous sections, holds for the interacting models, as well. 

To check the existence of the pure imaginary eigenvalues $\lambda_k$ and find the multiplicity of the null space, we employ the third quantization method for calculating the Liouvillian spectrum for $N$-fermion chains where $\mod(N,4)= 0$ subject to the periodic boundary conditions, as described in Sec.~\ref{sec:model}. As the conserved dynamics is quadratic and the jump operators are linear in fermionic basis the Liouville super-operator ($\mathcal{L}$) can be diagonalized in terms of $2N$ normal master modes acting on the Fock states of density operators~\cite{Prosen_2008}. The eigenvalues of the super-operator ($\lambda_k$) can be obtained directly from the spectrum of the \emph{shape} matrix, aka \emph{rapidities} ($\beta_i$) as
\begin{equation}~\label{eq:L spec from rapidity}
    \lambda_{\vec{v}} = - 2 \sum_{i = 1}^{2N} \beta_i v_i \, ,
\end{equation}
where $\vec{v}$ is a $2N$-long binary string. 

The whole Liouvillian spectrum therefore, can be exactly calculated by considering all $\vec{v}$ within $(1,4^N)$. Due to the linear growth of the shape matrix with $N$, in opposed to an exponential one, one can obtain detailed information about $\mathcal{L}$ without being limited to small $N$ chains hence, an equal treatment of the finite-sized systems and larger one approaching the thermodynamic limit (cf. Appendix~\ref{app:3rd quantization summary} for further details).

Figure~\ref{Fig1}(a)-(c) shows the Liouvillian spectrum ($\lambda_k$) of non-interacting Kitaev model in (\ref{eq:Hamiltonian}) for $\mu = 0, w= 1, \Delta = i$ at different chain lengths of N = 4, 12, and 100, respectively. For cases (b) and (c) the spetrum is zoomed in closed to the imaginary axis to highlight the slowly-varying modes. The red dots show the pure imaginary eigenvalues at $\lambda_\pm = \pm 2i$, and the green dot corresponds to the degenerate NESS at $\lambda_0 = 0$.

To examine the multi-stability and the long-time behavior of the system we looked at the two-point correlations and their time evolution. As the whole dynamics, including both the conservative and the dissipative part, is quadratic the state is Gaussian hence its first and second moments (two-point correlation functions) are sufficient to describe the system, fully. Using the Heisenberg picture we can derive the following equations of motion for the two-point correlation functions
\begin{widetext}
\begin{align}
    \frac{d}{dt} \braket{c_m c_n} & = \ii \left(w \braket{c_{m+1} c_n} + w \braket{c_m c_{n+1}} + w^* \braket{c_{m-1} c_n} + w^* \braket{c_m c_{n-1}} + 2\mu \braket{c_m c_n} - \Delta^* \braket{c_n c_{m+1}^\dagger} + \Delta^* \braket{c_n c_{m-1}^\dagger}\right) \\ \nonumber
    & + \ii \left(\Delta^* \braket{c_m c_{n+1}^\dagger} - \Delta^* \braket{c_m c_{n-1}^\dagger} \right) + 2 \sum_k \gamma_k \left(\left(c_k + c_k^\dagger \right) \left(c_m \delta_{nk} - c_n \delta_{mk} \right) \right)\, ,
\end{align}
\end{widetext}
\begin{widetext}
\begin{align}
    \frac{d}{dt} \braket{c_m^\dagger c_n} & = \ii \left(-w \braket{c_{m-1}^\dagger c_n} + w \braket{c_m^\dagger c_{n+1}} - w^* \braket{c_{m+1}^\dagger c_n} + w^* \braket{c_m^\dagger c_{n-1}} + \Delta \braket{c_{m-1} c_n} - \Delta \braket{c_{m+1} c_n}\right) \\ \nonumber
    & + \ii \left(-\Delta^* \braket{c_m^\dagger c_{n-1}^\dagger} + \Delta^* \braket{c_{n+1}^\dagger c_m^\dagger} \right) + 2 \sum_k \gamma_k \left(\left(c_n \delta_{mk} - c_m^\dagger \delta_{nk} \right) \left(c_k + c_k^\dagger \right)  \right)\, .
\end{align}
\end{widetext}
As can be seen correlations make a closed set of coupled non-linear equations that can be numerically solved for different randomized initial states. 

The time evolution of a local two-point correlation ($\braket{\hat{c}_1 \hat{c}_2}$) of such chains is presented in Fig.~\ref{Fig2} showcasing the emergence of both non-stationary steady states, aka dissipative time crystal, and the multistability. Since the long-time solutions in the multistable region depend on the initial state (cf. Eq.~\ref{eq:formalsolution}), the correlations equations of motion for the quadratic model are evolved for randomized initial states, two of them shown as red and blue lines in each panel. 

\begin{figure}[t]
    \includegraphics[width=0.9\linewidth]{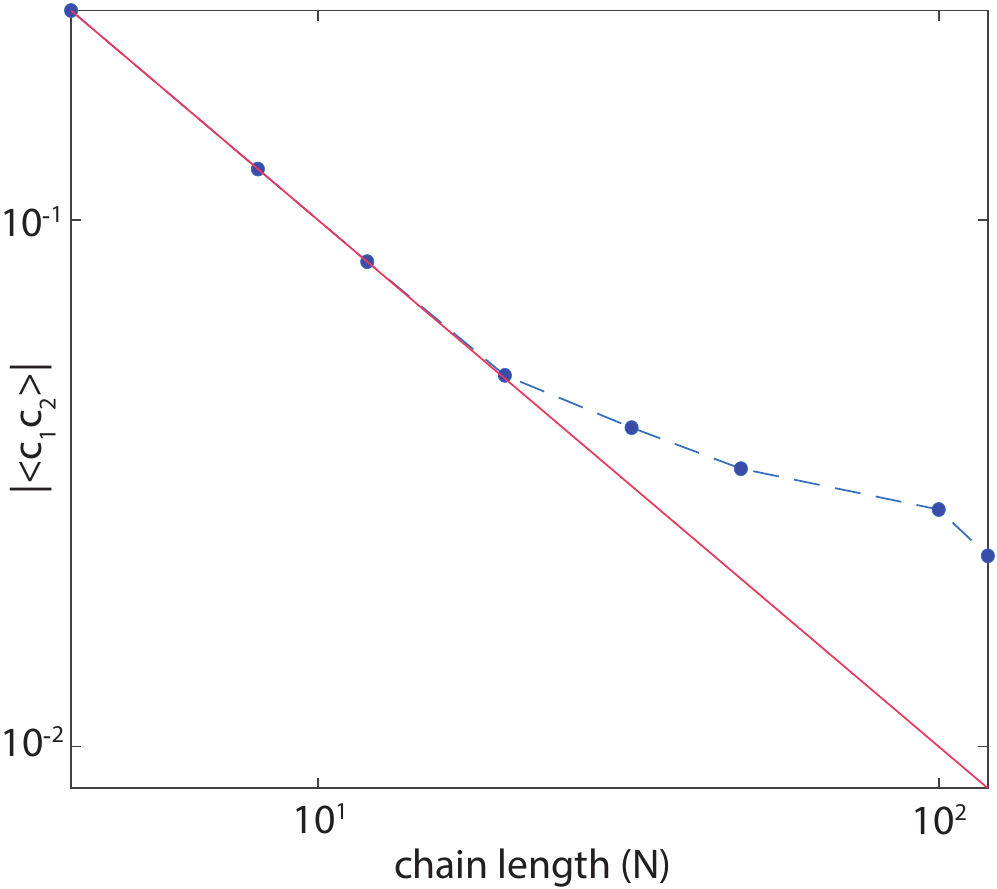}
    \centering
    \caption{~\label{Fig3} Local observable scaling vs. the chain length $N$ when the initial correlation in all cases is the same as $\braket{\hat{c}_i \hat{c}_j} = 1+i$ and $\braket{\hat{c}_i\hat{c}_i^\dagger} = 0$. The dots are the results of the moments equations of motion integration, the dashed line is guide to the eye, and the red line is $N^{-1}$ scaling for comparison.}
\end{figure}

To examine the scaling of the local observable with the system size ($N$) in Fig.~\ref{Fig3} we plot the long-time value of $\braket{\hat{c}_1 \hat{c}_2}$ as a function of the chain length.
%%%%%%%%%%%%%%%%%%%%%%%%%%%%%%%%%%%%%%%%%%%%%%%%%%%%%%%%%%%

%%%%%%%%%%%%%%%%%%%%%%%%%%%%%%%%%%%%%%%%%%%%%%%%%%%%%%%%%%%

The dots show the results of the numerical calculations when the initial correlations are chosen to be $\braket{ \hat{c}_n \hat{c}_m} = 1+\ii$ and $\braket{\hat{c}_n \hat{c}_n^\dagger} = 0$, i.e. having one particle on each site. As can be seen the correlation for smaller system sizes follows a power law behavior decaying as $N^{-1}$ (red line in Fig.~\ref{Fig3}) consistent with the lower bound prediction of Sec.~\ref{sec:model}. For larger $N$ more semi-local dynamical symmetries (semi-local finite frequency Goldstone modes) start emerging and the decay with $N$ slows, consistent with the results of Sec.~\ref{sec:TD and dynamical symmetries}. More specifically, 
\begin{equation}
        \ave{O}(t \to \infty) = \sum_k e^{t (\ii \omega_k+{\cal O}(1/N))} \sinner{O}{\rho_k} \sinner{\sigma_k}{\rho(0)}\,.
\end{equation}
with $\omega_k \in \mathbb{R}$ and the complex decay rate goes down as ${\cal O}(1/N))$. This implies both multi-stability and the persistent oscillations. 

\section{Conclusion}~\label{sec:conclution}
In this paper we have shown that for large classes of driven strongly interacting pairing models there exist spectrum generating algebras that are semi-local in the spin basis, which we therefore named semi-local dynamical symmetries. Physically, they correspond to particle excitations that are invisible to the interaction. They also imply non-local (quadratic) conservation laws, which are promoted to strong symmetries when the system is subjected to pair dephasing. Being quadratic, these operators directly imply memory of the initial condition, i.e. degenerate stationary states and bistability that decays with system size. Remarkably, this means that, unlike previously studied cases of bistability, here the bistability is induced by the quantum fluctuations and in the finite-size systems (beyond mean-field) rather than being destroyed by them. 

The system in the thermodynamic limit obtains emergent dynamical symmetries, which are finite-frequency and finite-momentum quasi-particles dressing the original dynamical symmetry excitations. Therefore, we call these emergent dynamical symmetries semi-local finite frequency and finite momentum Goldstone modes. For the dissipative system they imply strong symmetries. These Goldstone modes imply that, as we approach the thermodynamic limit, decay times of oscillations in the local observables diverge, but their amplitude goes to zero at least for initial states that have low-enough entanglement. Hence the system is a boundary time \emph{pseudo}-crystal according to the thermodynamic requirements of \cite{Fazio}. However, the oscillations are clean and periodic both for the isolated (closed $L_\mu=0$) and dissipative system, therefore our system is \emph{not} a dissipative time pseudo-crystal in the sense of the dissipative time crystals introduced in \cite{Buca_2019}, which would imply that dissipation is the one inducing periodic oscillations absent for the isolated system. 

To the best of our knowledge, this is the first exact and fully non-perturbative result on the long-debated problem of bistability and multistability in driven-dissipative many-body quantum systems.
Although we studied pairing fermionic models, the approach of thermodynamically emergent dynamical symmetries implying quasi-particles that are invisible to certain kinds of interactions is general and can be applied to both bosonic and spin systems. In future work, we plan to apply our approach to many-body spin and bosonic systems where multi-stability and persistent oscillations have been experimentally observed in the thermodynamic limit. Exploring the underlying connection between emergent collective behaviors and topology in such systems with quantum synchronization \cite{buca2022algebraic,quantumsynch,Roulet_2018,solanki2021role} are other interesting directions of the future studies. Importantly, our work provides an approach for proving presence of bistability in more general and widely studied quantum optical setups. 

\section*{Acknowledgments}
HA acknowledges the Purdue University Startup fund. BB acknowledges funding from the EPSRC programme grant EP/P009565/1, and the EPSRC National Quantum Technology Hub in Networked Quantum Information Technology (EP/M013243/1).

\bibliography{Ref}

\clearpage
\appendix

\section{Shape Matrix, Rapidities, and Liouvillian spectrum}~\label{app:3rd quantization summary}
As described in~\cite{Prosen_2008} the two parts of the super-operator, i.e. the conserved dynamics $\hat{\mathcal{L}}_H$ and the non-unitary parts $\hat{\mathcal{L}}_D$ has the following forms
\begin{equation}~\label{eq:LH}
    \mathcal{L}_H = -i 4 \sum_{j,k} \hat{c}_j^\dagger H_{jk} \hat{c}_k \, ,
\end{equation}
and
\begin{equation}~\label{eq:LD+}
    \mathcal{L}_D^+ = 2 \sum_{j,k=1}^{2N} \sum_{\mu=1}^N l_{\mu j} l^*_{\mu k}\left(2 \hat{c}_j^\dagger \hat{c}_k^\dagger - \hat{c}_j^\dagger \hat{c}_k - \hat{c}_k^\dagger \hat{c}_j \right) \, , 
\end{equation}
where $\hat{c}_i,\hat{c}_i^\dagger$ are the super-operator (a-fermion) annihilation and creation operators in the operator Fock space, respectively. Here, we focus on the $\mathcal{K}^+$, i.e. the even sup-space, only.

We define the $4N\times 1$-vector of a-fermionic operators as
\begin{equation}~\label{eq:a-fermion vector}
    \mathbf{\hat{C}} = \begin{pmatrix}
    \hat{c}_1 \\ \cdots \\ \hat{c}_{2N} \\ \hat{c}_1^\dagger \\ \cdots \\ \hat{c}_{2N}^\dagger \, .
    \end{pmatrix}
\end{equation}
With this definition we can write the Liouville super-operator $\hat{\mathcal{L}}^+ = \hat{\mathcal{L}}_H + \hat{\mathcal{L}}_D^+$ as
\begin{equation}~\label{eq:L+ quadratic form}
    \hat{\mathcal{L}}^+ = \mathbf{\hat{C}}^\dagger \mathbf{L}^+ \mathbf{\hat{C}} = \mathbf{\hat{C}}^\dagger 
    \begin{pmatrix}
    \mathbf{L}_{11}  & \mathbf{L}_{12} \\
    \mathbf{0}  & \mathbf{L}_{22}
    \end{pmatrix}
    \mathbf{\hat{C}}\, ,
\end{equation}
To write the non-unitary parts easier, we define a matrix with entries $M_{jk} = \sum_{\mu=1}^N l_{\mu j} l^*_{\mu k}$ hence, $\mathbf{M} = \mathbf{M}^\dagger$ is a Hermitian matrix.

Using these definitions we have
\begin{multline}~\label{eq:L matrix block diagonal}
    \begin{aligned}
    \mathbf{L}_{11}^{jk} & = -i2 H_{jk} - M_{jk} - M_{kj}  = -i2 H_{jk} - M_{jk} - M^'_{jk}\, ,  \\
    \mathbf{L}_{12}^{jk} & = 4 M_{jk}\, , \\
    \mathbf{L}_{22}^{jk} & = i2 H_{kj} + M_{kj} + M_{jk} = -i2 H_{jk} + M'_{jk} + M_{jk}  \, .
    \end{aligned}
\end{multline}
Considering the anti-symmetric properties of $\mathbf{H}$, it becomes apparent that $\mathbf{L}_{11} = -\mathbf{L}_{22}^\dagger$. Therefore, we have
\begin{equation}~\label{eq:quadratic L+}
    \hat{\mathcal{L}}^+ = \mathbf{\hat{C}}^\dagger 
    \begin{pmatrix}
    -\mathbf{L}^\dagger_{22}  & \mathbf{L}_{12} \\
    \mathbf{0}  & \mathbf{L}_{22}
    \end{pmatrix}
    \mathbf{\hat{C}}\, .
\end{equation}
The shape matrix $A$ defined in~\cite{Prosen_2008} is simply a rotation of this matrix hence, the eigenvalues of $A$ and $\mathbf{L}^+$ are the same. If $\eta_i, i \in \{1,2,\cdots,2N\}$ are the eigenvalues of $\mathbf{L}_{22}$ then the eigenvalues of $\mathbf{L}^+$ appear in pairs as $(\eta_i, -\eta_i^*)$. Also from the form of $\mathbf{L}_{22}$ it is clear that the eigenvalues appear in complex conjugate pairs as $(\gamma_i , \gamma_i^*), i \in \{1,2,\cdots,N\}$. Finally, one can conclude that the spectrum of the shape matrix $\mathbf{A}$ appear in quadruple of $(\xi, \xi^*, -\xi, -\xi^*)$. The rapidities are defined as the subset of the eigenvalues with positive real parts from which the full spectrum of $\mathcal{L}^+$ can be obtained using (\ref{eq:L spec from rapidity}). From this spectrum it becomes evident if there are any kernels, corresponding to multi-stability, or pure imaginary eigenvalues, corresponding to a non-stationary NESS.
\subsection{Dispersion of a chain with periodic boundary conditions}~\label{app:1D Kitaev chain PBC}
Let's consider an $N$-long chain with periodic boundary conditions (PBC), \textit{i.e.} $c_{m+N} = c_m$. One can use the following Fourier transformation to find the spectral form
\begin{equation}~\label{eq:Fourier series}
    \tilde{c}(k) = \frac{1}{\sqrt{N}} \sum_m e^{-imk} c_m ~, ~ c_m = \frac{1}{\sqrt{N}} \sum_k e^{i mk} \tilde{c}(k)\, .
\end{equation}
It is straightforward to see that the anti-commutator relations of the Fourier series has the following form
\begin{equation}~\label{eq:anti-commutator Fourier space}
    \{\tilde{c}(k) , \tilde{c}(k')\} = 0 ~ , ~ \{\tilde{c}(k) , \tilde{c}^\dagger(k')\} = \delta_{mn} \delta(k-k')\, .
\end{equation}
Replacing each term by its Fourier transform, we get the following spectral form 
\begin{widetext}~\label{eq:FT of the Hamiltonian}
    \begin{align}
        \tilde{H}(k) & = \sum_k -2|w| \cos{(k + \phi_w)} \tilde{c}^\dagger(k) \tilde{c}(k) -\frac{\mu}{2} \left(\tilde{c}^\dagger(k) \tilde{c}(k) - \tilde{c}(k) \tilde{c}^\dagger(k) \right) \\
        &+ \Delta e^{-ik} \tilde{c}(k) \tilde{c}(-k) + \Delta^* e^{-ik} \tilde{c}^\dagger(k) \tilde{c}^\dagger(-k) \\
        & = \begin{pmatrix}
        \tilde{c}^\dagger(k)  & \tilde{c}(-k)
        \end{pmatrix}
        \begin{pmatrix}
        -|w| \cos{(k + \phi_w)} - \frac{\mu}{2} && i \Delta^* \sin{k} \\
        -i\Delta \sin{k} && |w| \cos{(-k + \phi_w)} + \frac{\mu}{2}
        \end{pmatrix}
        \begin{pmatrix}
        \tilde{c}(k) \\ \tilde{c}^\dagger(-k)
        \end{pmatrix}\, .
    \end{align}
\end{widetext}

The choice of this spinor is useful since we can readily write the Fourier transform of Majorana fermions as a direct rotation
\begin{equation}~\label{eq:Majorana fermion FT}
    \begin{pmatrix}
    \tilde{w}_o(k) \\ \tilde{w}_e(k)
    \end{pmatrix} = 
    \begin{pmatrix}
    1 & 1 \\
    i & -i
    \end{pmatrix}
    \begin{pmatrix}
    \tilde{c}(k) \\ \tilde{c}^\dagger(-k)
    \end{pmatrix}\, ,
\end{equation}
where the $o,e$-superscripts refer to the odd and even Majorana fermions.

Substituting this back into the spectral Hamiltonian we can re-write it in terms of the Majorana fermions as 
\begin{widetext}
    \begin{align*}~\label{eq:spectral Hamiltonian Majorana}
    H_w & = \frac{1}{2} 
    \begin{pmatrix}
    |w|\sin{\phi_w} \sin{k} + \textrm{Im}(\Delta) \sin{k} &
    i \left(|w| \cos{\phi_w} \cos{k} + \frac{\mu}{2}\right) - \textrm{Re}(\Delta) \sin{k} \\
    -i \left(|w| \cos{\phi_w} \cos{k} + \frac{\mu}{2}\right) - \textrm{Re}(\Delta) \sin{k} & 
    |w|\sin{\phi_w} \sin{k} - \textrm{Im}(\Delta) \sin{k}
    \end{pmatrix}
\end{align*}
\end{widetext}

If the jump operators are identical for all fermionic sites as $L_j = \sqrt{g} \left(c_j + \delta c_j^\dagger \right)$, then $\mathbf{M}$ will read as follows
\begin{equation}
    \mathbf{M} = \frac{g}{4}
    \begin{pmatrix}
    |1+\delta|^2 & i \left(1 - |\delta|^2\right) - 2 \textrm{Im}(\delta) \\
    - i \left(1 - |\delta|^2 \right) - 2 \textrm{Im}(\delta) & |1-\delta|^2
    \end{pmatrix}
\end{equation}
Finally, we can use (\ref{eq:L matrix block diagonal}) to determine the rapidities from the eigenvalues of $\mathbf{L}_{22} = -i2\mathbf{H} + \mathbf{M} + \mathbf{M}'$. 
This leads to the following dispersion relation for rapidies as $\beta(k)$
\begin{equation}~\label{eq:PBC rapidities}
    \beta(k)  = \frac{g}{4}\left(|1+\delta|^2 + |1-\delta|^2 \right) - i |w| \sin{\phi_w} \sin{k} + \pm \frac{1}{2}\sqrt{\Lambda} \, ,
\end{equation}
where
\begin{widetext}~\label{eq:BPC imag of rapidities}
    \begin{align}
        \Lambda & = \ii 4g\left(|1+\delta|^2 - |1 - \delta|^2 \right) \textrm{Im}(\Delta) \sin{k} - 4 \left(|w| \cos{\phi_w} \cos{k} + \frac{\mu}{2}\right)^2 - 16 |\Delta|^2 \sin^2{k} \\
        & + \ii 16 g \textrm{Re}(\Delta) \textrm{Im}(\delta) \sin{k} + \frac{g^2}{4} \left(|1+\delta|^4 + |1-\delta|^4 \right)
        -\frac{g^2}{2} |1-\delta^2|^2 + 4 g^2 \textrm{Im}(\delta)^2\, .
    \end{align}
\end{widetext}

\end{document}